\documentclass[journal]{IEEEtran}

\usepackage[utf8]{inputenc}
\usepackage[T1]{fontenc}
\usepackage{graphicx}
\usepackage{booktabs}
\usepackage{array}
\usepackage[hyphens]{url}
\usepackage{tikz}
\usetikzlibrary{positioning,arrows.meta,calc}
\usepackage[hidelinks]{hyperref}

\newcommand{\code}[1]{\texttt{#1}}
\usepackage{xcolor}

\newcommand{\suppref}[1]{Appendix~#1 of the Supplementary Material}

\makeatletter
\renewenvironment{IEEEbiographynophoto}[1]{%
\if@IEEEbiographyTOCentrynotmade%
\setcounter{IEEEbiography}{-1}%
\refstepcounter{IEEEbiography}%
\addcontentsline{toc}{section}{Biographies}%
\global\@IEEEbiographyTOCentrynotmadefalse%
\fi%
\refstepcounter{IEEEbiography}%
\addcontentsline{toc}{subsection}{#1}%
\normalfont\footnotesize\interlinepenalty500%
\vskip 1.5\baselineskip plus 0.5\baselineskip minus 0\baselineskip%
\parskip=0pt\par%
\noindent\textbf{#1\ }\@IEEEgobbleleadPARNLSP}{\relax\par\normalfont}
\makeatother

\begin{document}

\title{A Multimodal Type~2 Diabetes Registry on DHIS2: A Design and Feasibility Study with the AI-READI Dataset}

\author{Saptarshi~Purkayastha,
        Aiswarya~Nandan~Perumbilly,
        Zhenan~Yin,
        and~Sherri~L.~Bucher%
\thanks{S.~Purkayastha, A.~Nandan Perumbilly and Z.~Yin are with the
Department of Biomedical Engineering and Informatics, Luddy School of
Informatics, Computing, and Engineering, Indiana University Indianapolis,
Indianapolis, IN, USA (corresponding author: S.~Purkayastha, e-mail:
saptpurk@iu.edu).}%
\thanks{S.~L.~Bucher is with the Department of Community and Global Health,
Richard M. Fairbanks School of Public Health, Indiana University
Indianapolis, Indianapolis, IN, USA.}%
}

\markboth{T2D Registry on DHIS2}%
{Purkayastha \MakeLowercase{\textit{et al.}}: A Multimodal Type~2 Diabetes Registry on DHIS2}

\maketitle

\begin{abstract}
Care for type~2 diabetes increasingly draws on continuous glucose monitors, wearable devices, home sensors, clinical records and retinal images, yet few open platforms have been shown to hold all of these for the same person. We asked three questions of DHIS2, the open-source health information system that ministries of health in about 80 countries run nationally: whether an unmodified Tracker program can represent these modalities together (RQ1), which design choices its row-per-event data model forces for dense sensor streams (RQ2), and what a bulk loader must do to load such data correctly (RQ3). We designed a registry of 27 program stages and 445 data elements, loaded the AI-READI dataset of 2,280 participants into it, and verified the result field by field against the source files. The program held every loaded modality in 8.09~million events. Dense streams were stored as hourly summaries aligned across modalities; electrocardiograms, diagnoses and 118,480 retinal images were stored as one event per record. The event model ruled out storing raw readings, and hourly summaries reduced the wearable and glucose volume about 26-fold. An audit of the loader found defects that shared one pattern: a failure that looked like success. We state five lessons for DHIS2 Tracker clients that address this pattern. Verification of all 64.9~million stored values found four defects, all missing, duplicated or artifactual records rather than wrong values, and none after their repair; an independent recomputation of the glucose summaries agreed on all 464,210 values compared. The paper offers implementation guidance for DHIS2 implementers and for users of AI-READI; it does not evaluate clinical use. Analytics at this scale, use by clinicians and live device feeds remain to be shown.
\end{abstract}

\begin{IEEEkeywords}
DHIS2, diabetes registry, continuous glucose monitoring, wearable devices, multimodal data integration, OMOP common data model, retinal imaging, health information systems, data quality.
\end{IEEEkeywords}

\section{Introduction}

\IEEEPARstart{T}{ype}~2 diabetes affects an estimated 589~million adults worldwide, most of them in low- and middle-income countries~\cite{genitsaridi2026}. Behavior and environment shape its course alongside care in the clinic~\cite{hillbriggs2021}, and exposure to fine particulate air pollution, indoors as well as outdoors, is one such influence~\cite{yang2020}. Many of these influences can now be measured. Continuous glucose monitors improve glucose control in type~2 diabetes~\cite{jancev2024}, are recommended for a growing share of patients~\cite{ada2026}, and are read through consensus measures such as the share of time spent between 70 and 180~mg/dL~\cite{battelino2019}. Wrist-worn devices record heart rate, sleep and activity. Retinal photographs can be graded for diabetic eye disease by software~\cite{gulshan2016} and carry signals about systemic disease~\cite{wagner2020}, and small sensors log the air in a room. Few reports describe a system that holds these streams for the same person over time in a form that clinicians and community health workers can use.

DHIS2 (District Health Information Software~2) is an open-source platform that grew out of the Health Information Systems Program, a network that began in South Africa in the 1990s~\cite{braa2004}. It was built for aggregate public health reporting and later gained a Tracker module for longitudinal, individual-level records~\cite{dhis2docs}. Ministries of health in about 80 countries run it as their national health information system~\cite{dhis2inaction, carnahan2023}, and Tracker has served as a national pregnancy registry~\cite{seebregts2018}, an electronic immunization registry~\cite{carnahan2023} and the record of hypertension and diabetes programs~\cite{nong2024, rtsl2024}. A platform with this reach is an attractive foundation for a registry that must outlive a single grant~\cite{paddo2024}. Whether its data model can hold dense sensor data, images and clinical records for thousands of participants, and what it takes to load such data correctly, has not been reported in the peer-reviewed literature (Section~\ref{sec:related}). We address three questions.

\begin{enumerate}
\item[RQ1] \emph{Representation.} Can an unmodified DHIS2 Tracker program represent continuous glucose, wearable, indoor environmental, electrocardiogram, diagnostic and retinal imaging data for the same participants alongside the clinical record?
\item[RQ2] \emph{Granularity and scale.} Which design choices does the platform's row-per-event data model force when the sources are dense sensor streams, and what does the result cost in events and in information?
\item[RQ3] \emph{Correctness.} What must a bulk loader do to put millions of records into such a system correctly through its web interface, and does the loaded registry agree with its source?
\end{enumerate}

We answer them in three steps, each applied to the AI-READI dataset~\cite{baxter2024, owsley2025}, a public, access-controlled collection of 2,280 participants across the spectrum of type~2 diabetes. First, we designed a registry data model that maps the modalities onto one Tracker program (RQ1, RQ2). Second, we loaded the dataset through the platform's web interface and audited the loader, deriving lessons from the defects found (RQ3). Third, we verified the loaded registry against the source files, field by field (RQ3). The contributions are a tested Tracker data model for dense consumer-device streams; five lessons for loading data into DHIS2 Tracker correctly; and a verification method with an account of the loaded registry, including a catalog of the data quality of AI-READI as an ingestion pipeline meets it. A preliminary account of the mapping model, covering 23 stages and 152 data elements, was presented as a conference abstract~\cite{perumbilly2026}. The pipeline code, program metadata and verification scripts are public\footnote{\url{https://github.com/iupui-soic/dhis2-diabetes-registry}}.

\section{Background}
\label{sec:background}

\subsection{The DHIS2 Tracker data model}

A Tracker \emph{program} enrolls \emph{tracked entities}, here participants, at an \emph{organisation unit} such as a clinic or study site. Each tracked entity carries \emph{attributes}, which are person-level values such as year of birth. An enrollment holds \emph{events}, each belonging to a \emph{program stage}; a \emph{repeatable} stage can hold any number of events per participant, each with its own \code{occurredAt} timestamp. Events carry \emph{data values}, one per \emph{data element}. Data elements are global objects with a value type and, optionally, an \emph{option set}, which is a controlled list of permitted values; one element can be attached to several stages. Images are stored as \emph{file resources} that an image-typed data value points to. Everything is exposed through a web interface, and bulk loading posts records to \code{/api/tracker}~\cite{dhis2docs}. Two properties of the model shape the design that follows. First, Tracker is row-oriented: each event is a database row, so the cost of a modality grows with its number of events rather than with the number of readings summarized in each event. Second, an event is identified by its 11-character UID~\cite{dhis2docs}. If the client does not supply one, the server generates it, and the server then has no way to recognize the same event when it is sent again.

\subsection{Related work}
\label{sec:related}

The first body of related work is the use of Tracker for individual-level records. The largest published deployments are in maternal and child health and immunization: South Africa's MomConnect program registered 1.77~million pregnancies in Tracker by 2017~\cite{seebregts2018}, and 43 countries used Tracker to manage vaccination during the coronavirus pandemic~\cite{carnahan2023}. For non-communicable disease, a Tracker application screened 3,485 older adults for hypertension and diabetes in Vietnam~\cite{nong2024}, and a standard Tracker package for hypertension and diabetes care is maintained with Resolve to Save Lives~\cite{rtsl2024}. All of these hold records that a health worker enters at an encounter. Device data have reached DHIS2 in one domain, the vaccine cold chain: an OpenFn template imports Fridge-tag temperature loggers into Tracker with readings summarized into daily events~\cite{openfnberlinger}, which is the same choice of event unit made here at hourly resolution. These integrations monitor equipment; none records data about a person. Images have been placed in Tracker by script, for example from obstetric ultrasound in Guatemala~\cite{guatemala2022}. All of these integrations are described in community reports and have not been published in peer-reviewed form.

The second is DHIS2 as a research data platform. DHIS2 data are used mostly for planning and performance monitoring~\cite{dehnavieh2019}. The closest precedent to this work is a wound care research registry that joined 277 patients from six studies to area-level socioeconomic data~\cite{paddo2024}; it held no sensor or imaging data.

The third is the set of platforms that research cohorts use for multimodal data. REDCap is the conventional tool for research data capture~\cite{harris2009}; RADAR-base collects data from wearables and phones through streaming storage~\cite{ranjan2019}; and the MIMIC critical care database keeps high-frequency waveforms in a store linked to the relational record~\cite{johnson2016}. None of these is deployed as a national health information system. For glucose monitors, the iCoDE consensus sets standards for bringing monitoring data into the electronic health record~\cite{icode2022, xu2022icode}. Integration can range from static reports, through discrete summary statistics such as time in range, to continuous structured data; the hourly summaries of this registry sit between the last two. AI-READI's own consortium has mapped its survey data into the OMOP common data model~\cite{ohdsiaireadi2024}. We searched the peer-reviewed literature and the DHIS2 community archives through October 2026 for sensor, imaging or OMOP data held in DHIS2 and for AI-READI loaded into a health information system; the community reports above are the closest precedents.

\subsection{The AI-READI dataset}

AI-READI is a flagship dataset of the Bridge2AI program of the National Institutes of Health~\cite{baxter2024}, enrolling adults aged forty and over at three sites~\cite{owsley2025}. Release~3.0.0~\cite{aireadi2025, aireadidocs} contains, for 2,280 participants, a Garmin Vivosmart~5 wearable export in the Open mHealth format~\cite{estrin2010} (heart rate, respiratory rate, blood oxygen saturation, stress, sleep, activity, calories); a Dexcom~G6 continuous glucose monitor export; an indoor environmental sensor file at five-second resolution (particulate matter in four size fractions, humidity, temperature, volatile organic compounds, nitrogen oxides); 12-lead electrocardiograms in WFDB format~\cite{goldberger2000}; retinal photography and optical coherence tomography angiography (OCTA) in DICOM~\cite{dicom}; clinical data in the OMOP common data model~\cite{hripcsak2015}; and a participant table. The release also holds structural optical coherence tomography and fluorescence lifetime imaging, which were not in the copy available to this project. The public release withholds participant sex, race and ethnicity~\cite{aireadi2025}: no OMOP gender, race or ethnicity concept is set, and the DICOM Patient Sex attribute is a placeholder. Where this paper notes that the registry holds no sex attribute, the attribute is missing because the public release withholds it. The dataset may not be redistributed~\cite{aireadilicense}.

\section{Methods}
\label{sec:methods}

\subsection{Step 1: Registry design}
\label{sec:design}

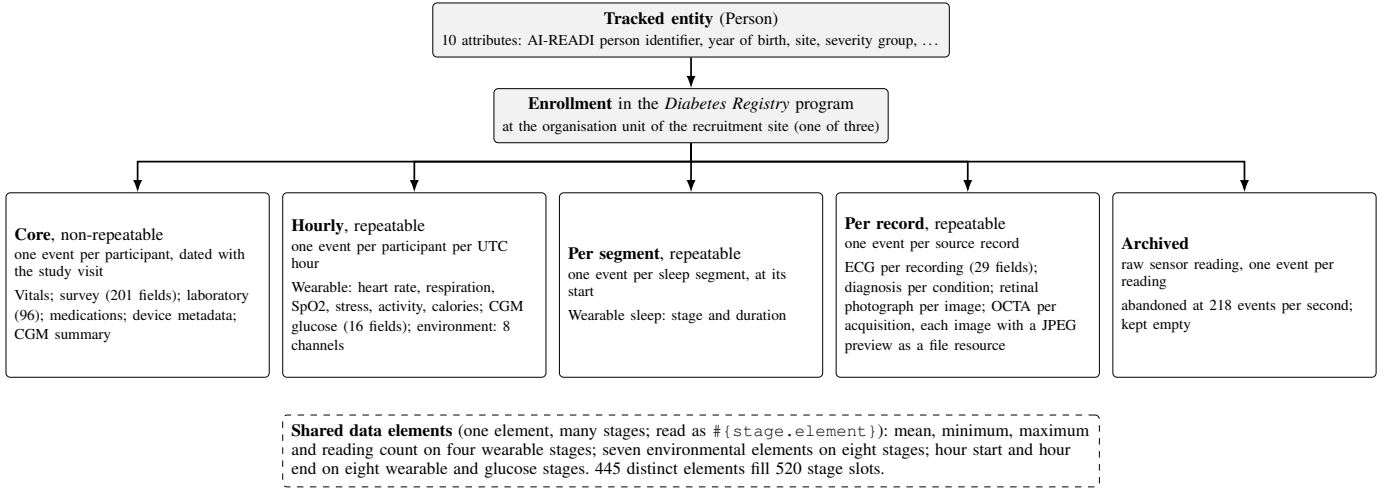
\begin{figure*}[!t]
\centering
\resizebox{\textwidth}{!}{%
\begin{tikzpicture}[
  font=\footnotesize,
  ent/.style={draw, rounded corners=2pt, align=center, inner sep=4pt, fill=black!5},
  grp/.style={draw, rounded corners=2pt, align=left, inner sep=4pt, text width=40mm, minimum height=30mm, anchor=north},
  shr/.style={draw, dashed, rounded corners=2pt, align=left, inner sep=4pt},
  arr/.style={-{Latex[length=2mm]}, thick},
]
\node[ent] (te) {\textbf{Tracked entity} (Person)\\[1pt]\scriptsize 10 attributes: AI-READI person identifier, year of birth, site, severity group, \dots};
\node[ent, below=5mm of te] (en) {\textbf{Enrollment} in the \emph{Diabetes Registry} program\\[1pt]\scriptsize at the organisation unit of the recruitment site (one of three)};
\draw[arr] (te) -- (en);
\coordinate (row) at ([yshift=-8mm]en.south);
\node[grp] (core) at ($(row)+(-90mm,0)$) {\textbf{Core}, non-repeatable\\[1pt]\scriptsize one event per participant, dated with the study visit\\[2pt]Vitals; survey (201 fields); laboratory (96); medications; device metadata; CGM summary};
\node[grp] (hour) at ($(row)+(-45mm,0)$) {\textbf{Hourly}, repeatable\\[1pt]\scriptsize one event per participant per UTC hour\\[2pt]Wearable: heart rate, respiration, SpO2, stress, activity, calories; CGM glucose (16 fields); environment: 8 channels};
\node[grp] (seg) at ($(row)+(0mm,0)$) {\textbf{Per segment}, repeatable\\[1pt]\scriptsize one event per sleep segment, at its start\\[2pt]Wearable sleep: stage and duration};
\node[grp] (rec) at ($(row)+(45mm,0)$) {\textbf{Per record}, repeatable\\[1pt]\scriptsize one event per source record\\[2pt]ECG per recording (29 fields); diagnosis per condition; retinal photograph per image; OCTA per acquisition, each image with a JPEG preview as a file resource};
\node[grp] (arch) at ($(row)+(90mm,0)$) {\textbf{Archived}\\[1pt]\scriptsize raw sensor reading, one event per reading\\[2pt]abandoned at 218 events per second; kept empty};
\foreach \n in {core,hour,seg,rec,arch} {\draw[arr] (en.south) -- ++(0,-3mm) -| (\n.north);}
\node[shr, below=6mm of hour.south west, anchor=north west, text width=130mm] (shared) {\textbf{Shared data elements} (one element, many stages; read as \code{\#\{stage.element\}}):\ mean, minimum, maximum and reading count on four wearable stages; seven environmental elements on eight stages; hour start and hour end on eight wearable and glucose stages. 445 distinct elements fill 520 stage slots.};
\end{tikzpicture}}
\caption{Data model of the registry: one tracked entity per participant, one enrollment at the recruitment site, and 27 program stages grouped by the unit of their events. Counts in parentheses are fields per stage.}
\label{fig:schema}
\end{figure*}

Each participant is one tracked entity, enrolled once in a program named \emph{Diabetes Registry} at the organisation unit of their recruitment site (Fig.~\ref{fig:schema}). The AI-READI person identifier is the only attribute shared with the source; nine others carry year of birth, diabetes severity group and similar person-level values. Six non-repeatable core stages hold one event per participant: vitals, a lifestyle survey, laboratory results, medications, device metadata and a glucose monitoring summary. The other stages are repeatable (Table~\ref{tab:stages}).

\begin{table}[!t]
\renewcommand{\arraystretch}{1.15}
\caption{Program Stages of the Registry on 30 October 2026}
\label{tab:stages}
\centering
\footnotesize
\setlength{\tabcolsep}{4pt}
\begin{tabular}{@{}l l r r@{}}
\toprule
Stage & Unit of event & Fields & Events \\
\midrule
\multicolumn{4}{@{}l}{\emph{Core (one per participant)}} \\
Vitals \& Measurements & visit & 9 & 2,280 \\
Health \& Lifestyle Survey & visit & 201 & 2,280 \\
Clinical Laboratory \& Assessment & visit & 96 & 2,277 \\
Medications - Current & visit & 10 & 2,203 \\
Device \& Sensor Metadata & visit & 6 & 2,154 \\
CGM Summary & visit & 3 & 2,245 \\
\midrule
\multicolumn{4}{@{}l}{\emph{Wearable (Garmin Vivosmart 5)}} \\
Wearable - Heart Rate & hour & 13 & 485,951 \\
Wearable - Respiratory Rate & hour & 13 & 487,770 \\
Wearable - SpO2 & hour & 13 & 73,001 \\
Wearable - Stress & hour & 6 & 479,077 \\
Wearable - Activity & hour & 4 & 870,911 \\
Wearable - Calories & hour & 4 & 325,201 \\
Wearable - Sleep & segment & 4 & 395,999 \\
\midrule
CGM - Glucose (Dexcom G6) & hour & 16 & 519,617 \\
\midrule
\multicolumn{4}{@{}l}{\emph{Environment (indoor sensor)}} \\
Environment - PM1 & hour & 7 & 538,179 \\
Environment - PM2.5 & hour & 7 & 538,179 \\
Environment - PM4 & hour & 7 & 538,179 \\
Environment - PM10 & hour & 7 & 538,179 \\
Environment - Humidity & hour & 11 & 538,179 \\
Environment - Temperature & hour & 7 & 538,179 \\
Environment - VOC & hour & 7 & 538,179 \\
Environment - NOx & hour & 7 & 538,169 \\
\midrule
Cardiac - 12-Lead ECG & recording & 29 & 2,257 \\
Diagnosis History (OMOP) & condition & 3 & 12,375 \\
Retinal Photography & image & 11 & 93,920 \\
Retinal OCTA & acquisition & 16 & 24,560 \\
\midrule
Archived raw reading stage & reading & 3 & (unused) \\
\midrule
\textbf{Total} & & 520 slots & \textbf{8,089,500} \\
\bottomrule
\end{tabular}
\vspace{1ex}
\begin{flushleft}
\scriptsize Stage names are reproduced as they appear on the server. Fields counts the data elements on each stage; the 520 slots hold 445 distinct data elements because some are shared between stages (Section~\ref{sec:granularity}). CGM is the continuous glucose monitor; SpO2 is blood oxygen saturation; PM1 to PM10 are particulate matter size fractions; VOC is volatile organic compounds; NOx is nitrogen oxides; ECG is electrocardiogram; OCTA is optical coherence tomography angiography.
\end{flushleft}
\end{table}

\emph{Event unit.} \label{sec:granularity}The central design decision is the unit of a sensor event. A Garmin export carries a reading every minute or two per stream, and an environmental file carries about 720 readings per hour per channel. Storing each raw reading as an event was tried first and abandoned; the stage remains in the program, archived and empty. Every dense stream is instead summarized into one event per participant, stage and hour of Coordinated Universal Time (UTC), with the hour start as \code{occurredAt}. This bounds the event count while keeping the resolution at which the consensus glucose measures are defined~\cite{battelino2019}, and it places glucose, heart rate and indoor air for one person in one hour in events with the same timestamp. Heart rate, respiratory rate, blood oxygen and stress hours hold mean, minimum, maximum and reading count. Glucose hours add the standard deviation, the three consensus range percentages~\cite{battelino2019, ada2026}, counts of the sensor's out-of-range \emph{markers} (the strings \code{High} and \code{Low} that the export writes above 400 or below 40~mg/dL, which are kept in the range denominator), and counts and times of readings above 180 or below 70~mg/dL. Standard deviations are population standard deviations and are left empty for an hour with fewer than two readings, as are the glucose mean, minimum and maximum of an hour that holds only markers; the stored reading count distinguishes these hours from hours without data, which have no event. Activity segments of up to 167 minutes are prorated across hours by time overlap, which preserves totals but smooths bursts of walking. Sleep is kept as segments. Environmental channels keep the sensor's units, with volatile organic compounds and nitrogen oxides as index points~\cite{sen5x}.

\emph{Shared elements.} Generic elements serve several stages: four summary elements on four wearable stages, seven environmental elements on eight stages, and hour start and end on eight stages, so 445 elements fill 520 stage slots. A shared element has one value type and one aggregation type everywhere, so sharing is confined to statistics of the same kind. A shared value is interpretable only together with its stage, and DHIS2 always carries the stage in the reference: a program indicator names a value as \code{\#\{stage.element\}}~\cite{dhis2docs}.

\emph{Time base.} The three sites lie in two time zones that observe daylight saving time. Storing UTC aligns modalities exactly and avoids the repeated and skipped hours of local time. The cost is that the Capture application shows UTC clock times and that a DHIS2 calendar day is a UTC day. Local hours remain recoverable without loss, because every event is recorded at its site's organisation unit and every site is a whole number of hours from UTC.

\emph{Flags and sufficiency.} Where a reference range exists, hourly stages count readings outside it and record their times and a status: heart rate outside 60 to 100 beats per minute, respiratory rate outside 12 to 20, blood oxygen below 95 and 90~percent, humidity outside 30 to 50~percent. These flags mark exceedance of a reference range and should not be read as clinical findings, because resting ranges are exceeded in the ordinary course of a day. Particulate matter carries no hourly flag because its guideline values are 24-hour means~\cite{whoaqg2021}, and the index channels carry none because no concentration guideline applies to them; temperature flag fields were specified but never attached. The electrocardiogram stage classifies heart rate, PR interval and QRS duration. It does not classify the corrected QT interval, whose reference ranges are sex-specific~\cite{rautaharju2009}. A Data Sufficiency status marks each heart rate, respiratory rate and oxygen hour as having no valid readings, one or two, or three or more, after removal of device placeholder values (\emph{sentinels}) and non-finite values. The threshold is a deliberately low project convention, and the stored count is the measure of density.

\emph{Per-record stages.} Electrocardiograms, diagnoses and images map to one event per source record: 29 fields parsed from each electrocardiogram header, including the first three machine findings; code, label and date for each OMOP condition; and, for each retinal photograph or OCTA acquisition, the manifest metadata, the DICOM SOP Instance identifier and an 800-pixel JPEG preview held as a file resource. For OCTA the preview is the device's en-face image of the superficial vascular plexus~\cite{sun2021}; full volumes are referenced by path.

\subsection{Step 2: Loading}
\label{sec:reliability}

\begin{figure*}[!t]
\centering
\resizebox{\textwidth}{!}{%
\begin{tikzpicture}[
  font=\footnotesize,
  node distance=7mm and 5mm,
  box/.style={draw, rounded corners=2pt, align=center, minimum height=19mm,
              inner sep=3pt, text width=27mm},
  side/.style={draw, dashed, rounded corners=2pt, align=center,
               minimum height=15mm, inner sep=3pt, text width=27mm},
  arr/.style={-{Latex[length=2mm]}, thick},
  lbl/.style={fill=white, inner sep=1.5pt, font=\scriptsize},
]
\node[box] (src) {\textbf{AI-READI source}\\[1pt]\scriptsize manifests, Open mHealth JSON, CSV, WFDB, DICOM, OMOP};
\node[box, right=of src] (extract) {\textbf{Extract and summarize}\\[1pt]\scriptsize UTC hour buckets, missing-value and sentinel filters, DICOM to JPEG};
\node[box, right=of extract] (build) {\textbf{Build events}\\[1pt]\scriptsize derived identifiers, option codes, finite-value guard};
\node[box, right=of build] (recon) {\textbf{Reconcile}\\[1pt]\scriptsize match existing events by identity, reuse their identifiers};
\node[box, right=of recon] (write) {\textbf{Write and verify}\\[1pt]\scriptsize create-and-update strategy, per-type import counts};
\node[box, right=of write, fill=black!5] (dhis) {\textbf{DHIS2 2.44-SNAPSHOT}\\[1pt]\scriptsize Tracker program, 27 stages, 445 data elements};

\node[side, below=of build] (meta) {\textbf{Metadata registry}\\[1pt]\scriptsize identifiers resolved by name, cached locally};
\node[side, below=of write] (ckpt) {\textbf{Checkpoint}\\[1pt]\scriptsize done / failed / skipped, atomic writes};
\node[side, below=of dhis] (verify) {\textbf{Verification}\\[1pt]\scriptsize read-only, every stored field against the source};

\draw[arr] (src) -- (extract);
\draw[arr] (extract) -- (build);
\draw[arr] (build) -- (recon);
\draw[arr] (recon) -- (write);
\draw[arr] (write) -- (dhis);
\draw[arr] (meta) -- (build);
\draw[arr] (meta.east) -- ++(2.5mm,0) |- ([yshift=-3mm]recon.south) -- (recon.south);
\draw[arr] (write) -- (ckpt);
\draw[arr] (dhis) -- (verify);
\draw[arr] (verify) -- (ckpt);
\coordinate (feed) at ([yshift=5mm]dhis.north);
\draw[arr, dashed] (dhis.north) -- (feed) -| (recon.north);
\node[lbl] at ($(feed)!0.5!(recon.north |- feed)$) {existing events of the participant, fetched once per stage};
\end{tikzpicture}}
\caption{Pipeline for one modality. Step~1 scripts populate the metadata registry. Step~2 scripts run the horizontal flow once per participant. Verification reads the registry back against the source files.}
\label{fig:arch}
\end{figure*}
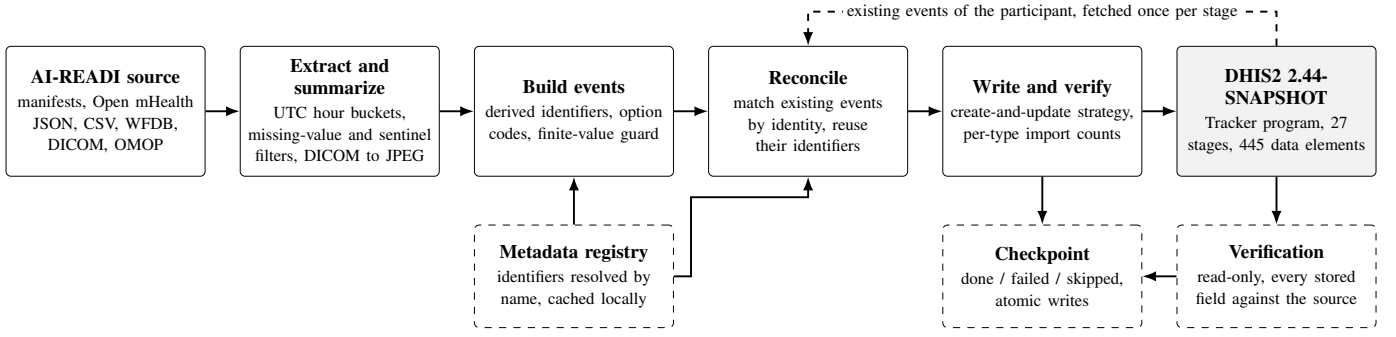

Each modality is loaded in two steps (Fig.~\ref{fig:arch}). A first script creates the metadata idempotently. A second script builds events from the source, reconciles them with events already on the server, posts them in sequential verified batches and records the outcome in a checkpoint. The program, with its six interactively configured core stages, is exported from the server as one metadata package so that it can be recreated in a single import.

The first version of the loader was audited by an independent information technology reviewer at Indiana University, who graded findings as \emph{critical} if they could lose data, \emph{high} if they could leave a stored value or attribute differing from its source, and \emph{medium} if they could duplicate data. The defects shared one pattern: in each case, a failure could not be told apart from success. The controls that close them are established practice in data engineering and distributed systems~\cite{yuan2014}. The specific contribution here is to show how each defect arises in a DHIS2 Tracker client, through interface behaviors that are not documented prominently. We therefore state the controls as five lessons for Tracker clients (Table~\ref{tab:principles}; the audit findings are listed in \suppref{B}).

\begin{table}[!t]
\renewcommand{\arraystretch}{1.2}
\caption{Lessons for DHIS2 Tracker Clients, the Defect Class Each Closes, and the Evidence}
\label{tab:principles}
\centering
\scriptsize
\setlength{\tabcolsep}{3pt}
\begin{tabular}{@{}p{2.1cm} p{2.9cm} p{3.2cm}@{}}
\toprule
Lesson & Defect class closed & Evidence \\
\midrule
Client-assigned identifiers & Retries, resumed runs and re-runs duplicate every record; records loaded before client identity existed are duplicated on re-run & Re-run over a participant with 3,798 existing events updated every one and created none \\
Writes verified from the response & Success status with failure in the body; aggregate statistics that hide a wrong object count; a stage built on a metadata reference that does not exist & Loader stops with an error on any batch whose per-type created plus updated count differs from the batch size; verification found no source hour without an event (Table~\ref{tab:verify}) \\
Checkpoints that record outcomes & A rejected participant marked done and skipped forever; an interrupted run truncating its own checkpoint & Checkpoints distinguish done, failed and skipped; failed and skipped work is retried on every run \\
Metadata by name, options by code & Hand-pasted identifiers; a stage duplicated on every re-run because its name on the server differed by one character; option values written as display names & Step~1 scripts re-run without creating objects; pointing the pipeline at another instance fails rather than writing into the wrong stage \\
Type and time checks at input & Not-a-number stored as the text \code{nan}; hours compared as formatted strings; values cleared or kept by an undocumented update semantics & Finiteness guard on every numeric value; UTC confirmed by diurnal check; merge-on-update verified on a sample of 300 updated events \\
\bottomrule
\end{tabular}
\end{table}

\emph{Client-assigned identifiers.} Every record carries a UID derived by hashing its source identity, and every write uses the create-and-update strategy, so a retry or re-run updates the record instead of duplicating it. Events loaded before identifiers were derived are matched to their source by identity before each write. \emph{Writes verified from the response.} A tracker import can report success with an error in its body, and its top-level counts sum every object type; the loader reads the per-type counts and stops with an error unless they equal the batch size. \emph{Checkpoints that record outcomes.} A participant is marked done only after a confirmed write, and failed or skipped work is retried. \emph{Metadata by name, options by code.} Identifiers are resolved from the live program by name, tolerant of dash characters, and option values are written as the server's codes. \emph{Type and time checks at input.} Non-finite values are dropped before aggregation, hours are compared as instants, and an update always carries the event's existing values, because the server's merge behavior is version-specific.

\subsection{Step 3: Verification}
\label{sec:verification}

Verification issues only read requests and compares each stored value with the value recomputed from the source. This is the strongest of the methods cataloged for assessing clinical data quality~\cite{weiskopf2013} and corresponds to the source data verification used in clinical trials~\cite{tudursmith2012}; here it is applied to every participant, every event and every stored field of the 26 stages in use. Hourly values are recomputed with the importer's own aggregation code, which establishes that the registry holds what that code computed but cannot detect an error in the code itself. The glucose aggregation was therefore also recomputed independently for a random sample of 200 participants: a separately written parser read the raw export, readings were bucketed by hour with pandas, and the metrics were computed with the iglu package~\cite{broll2021}. Per-record values are recomputed by code written apart from the importers. Every comparison reports two further counts, registry events with no source counterpart and source records with no event, so that a misalignment cannot be hidden by a perfect match over an empty set of values. A pair that is empty on both sides counts as agreement only where a rule on the source requires it, such as the standard deviation of a single reading or a field the source record does not carry, and such pairs are reported separately. The matching rules are detailed in \suppref{C}.

\subsection{Evaluation setting}
\label{sec:setting}

The registry runs on one virtual machine with six processor cores and 15~GiB of memory, under a pre-release development build of DHIS2 (2.44-SNAPSHOT) and PostgreSQL~16.9 at its default settings (\suppref{A}). Every timing reported here was taken on that untuned installation from a single client. Scale was assessed by the event counts of the loaded registry, by a pilot that loaded raw readings as events in an earlier phase~\cite{perumbilly2026}, by a cohort query issued through the transactional tracker interface, and by one attempt to generate the analytics tables that DHIS2 dashboards and program indicators read~\cite{dhis2analytics, dhis2perf}. Coverage and sufficiency were computed from read-only queries and from the source files.

\section{Results}
\label{sec:results}

\subsection{RQ1: Representation}
\label{sec:rq1}

\begin{figure*}[!t]
\centering
\begin{tabular}{@{}c@{\hspace{6mm}}c@{}}
\includegraphics[height=6.6cm]{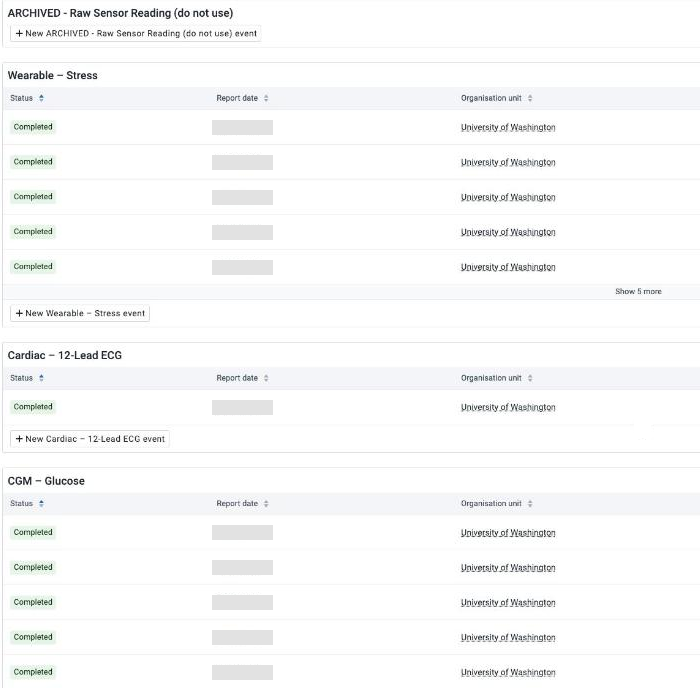} & \includegraphics[height=6.6cm]{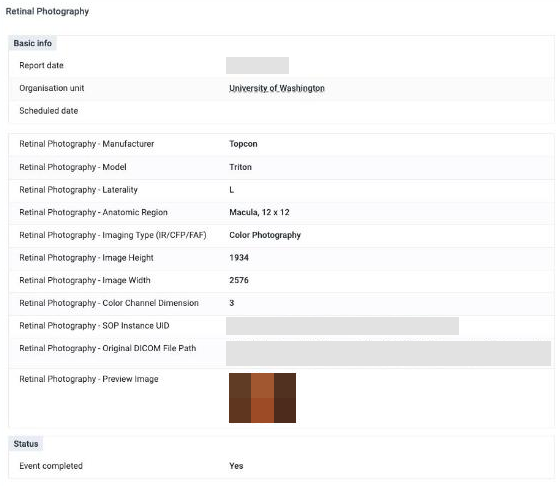} \\[1mm]
\footnotesize (a) & \footnotesize (b)
\end{tabular}
\caption{One participant's record in the unmodified Capture application. (a) The enrollment dashboard lists each stage's events by date alone, five at a time, without the hour or any value; the archived raw-reading stage is empty. (b) A retinal photography event with its manifest fields and its JPEG preview, held as a file resource. Dates, per-participant event counts, the DICOM identifier and the file path are masked, and the preview is pixelated.}
\label{fig:capture}
\end{figure*}

The unmodified Tracker program held every loaded modality for the same participants: 2,280 tracked entities with 10 attributes, 27 stages (26 in use) and 445 data elements, with 8,089,500 events in all (Table~\ref{tab:stages}). Repeatable stages carried up to 870,911 events each; option sets enforced controlled vocabularies; and file resources held 118,440 previews for 118,480 images, the 40 without one being OCTA acquisitions whose source has no superficial-plexus image. The Capture application displayed every stage and value without custom code (Fig.~\ref{fig:capture}). For a dense stream, however, display falls short of use: the default enrollment dashboard lists a stage's events by date alone, without the hour or any value, five at a time, so a participant's several hundred hourly events offer no practical view of the stream.

\subsection{RQ2: Granularity and scale}
\label{sec:rq2}

\begin{table*}[!t]
\renewcommand{\arraystretch}{1.15}
\caption{Verification of Every Stored Field Against the Source Files After Repair, by Stage, 30 October 2026}
\label{tab:verify}
\centering
\footnotesize
\setlength{\tabcolsep}{4pt}
\newcommand{\hd}[1]{\begin{tabular}[b]{@{}r@{}}#1\end{tabular}}
\begin{tabular}{@{}l r r r r r r r r@{}}
\toprule
Stage & Fields & Events & \hd{Values\\compared} & Matched & \hd{Empty,\\expected} & \hd{Mis-\\matched} & \hd{No\\source} & \hd{No\\event} \\
\midrule
\multicolumn{9}{@{}l}{\emph{Hourly and segment stages, recomputed with the importer's aggregation code}} \\
CGM - Glucose & 16/16 & 519,617 & 8,313,872 & 7,365,732 & 948,140 & 0 & 0 & 0 \\
Wearable - Heart Rate & 13/13 & 485,951 & 6,317,363 & 5,633,805 & 683,558 & 0 & 0 & 0 \\
Wearable - Respiratory Rate & 13/13 & 487,770 & 6,341,010 & 5,538,665 & 802,345 & 0 & 0 & 0 \\
Wearable - SpO2 & 13/13 & 73,001 & 949,013 & 912,155 & 36,858 & 0 & 0 & 0 \\
Wearable - Stress & 6/6 & 479,077 & 2,874,462 & 2,874,462 & 0 & 0 & 0 & 0 \\
Wearable - Activity & 4/4 & 870,911 & 3,483,644 & 3,483,644 & 0 & 0 & 0 & 0 \\
Wearable - Calories & 4/4 & 325,201 & 1,300,804 & 1,300,804 & 0 & 0 & 0 & 0 \\
Wearable - Sleep (segments) & 4/4 & 395,999 & 1,583,996 & 1,583,996 & 0 & 0 & 0 & 0 \\
Environment - Humidity & 11/11 & 538,179 & 5,919,969 & 5,023,521 & 896,448 & 0 & 0 & 0 \\
Environment - 7 other channels & 7/7 & 3,767,243 & 26,370,701 & 26,370,629 & 72 & 0 & 0 & 0 \\
\multicolumn{9}{@{}l}{\emph{Per-record stages, recomputed by separately written code}} \\
Diagnosis History & 3/3 & 12,375 & 24,750 & 24,750 & 0 & 0 & 0 & 0 \\
Retinal Photography & 11/11 & 93,920 & 939,200 & 939,200 & 0 & 0 & 0 & 0 \\
Retinal OCTA & 16/16 & 24,560 & 368,400 & 368,360 & 40 & 0 & 0 & 0 \\
Cardiac - 12-Lead ECG & 29/29 & 2,257 & 63,196 & 58,767 & 4,429 & 0 & 0 & 0 \\
\midrule
\textbf{All 20 stages} & & 8,076,061 & 64,850,380 & 61,478,490 & 3,371,890 & 0 & 0 & 0 \\
\midrule
CGM - Glucose, recomputed with iglu (200 participants) & 10/16 & 46,421 & 464,210 & 463,487 & 723 & 0 & 0 & 0 \\
Six core stages (earlier phase, separate script) & all & 13,439 & 82,302 & 82,302 & & 0 & & \\
\bottomrule
\end{tabular}
\vspace{1ex}
\begin{flushleft}
\scriptsize Every participant and every event of each stage was compared. Fields gives the fields compared out of those the stage stores; a field used to match an event to its source record (header path, SOP Instance identifier, condition code, sleep start) is counted as compared, because a wrong value leaves the event unmatched. Empty, expected counts pairs empty on both sides where a rule on the source requires it (Section~\ref{sec:verification}); most are lists of flagged times for hours in which no reading was flagged. No source counts registry events with no counterpart in the source; no event counts source hours or records with no registry event. Values are compared with a tolerance of 0.01. The run followed the four repairs described in the text. The iglu and core-stage rows use other methods and are not part of the totals.
\end{flushleft}
\end{table*}

Loading raw readings as events ran at about 218 events per second in the pilot, which extrapolates to five to six days for the roughly 100~million raw readings of the cohort. Hourly summaries reduced the wearable and glucose volume about 26-fold, to 3.8~million events that imported in under 45 minutes. The completed registry holds about 3,500 events per participant; the eight environmental stages are the largest component, at 4.31~million events. Through the transactional interface, assembling a cohort of 2,205 participants with a retinal preview and at least seven days of glucose took 184~seconds, as two requests per participant at about 40~ms each. Analytics table generation built the tracked-entity analytics tables, including one of every event in the registry, in about 15~minutes. It then failed while preparing the event analytics table, reporting one failed item but no cause through the interface, so the analytics interface could not be measured.

\emph{Coverage and sufficiency.}\label{sec:coverage} Over 99~percent of heart rate and respiratory rate hours and 89.5~percent of oxygen hours met the sufficiency threshold (Table~\ref{tab:coverage}). The median hour held 45 to 53 readings, however, so the reading count is the more informative measure. Three quarters of the cohort (1,709 of 2,280) yielded at least seven days of glucose, wearable and environmental data together without contact from a health worker. Most participants (85~percent) wore the glucose sensor for 11 days, and only 2 wore it for 14 days or more, which is short of the 14 days the consensus recommends~\cite{battelino2019, kok2026}. The exceedance flags fired in most hours: a heart rate reading outside the resting range occurred in 58.5~percent of hours and an oxygen reading below 95~percent in 95.2~percent. Counting days in site-local rather than UTC dates changed the number of days with glucose data for 191 of 2,245 participants and changed the seven-day classification for 5. As an example of a program-level count, glucose hours with less than 70~percent of readings in range made up 18.9, 13.5 and 13.2~percent of hours at the three sites; the count was obtained with six filtered tracker requests in 27~seconds.

\begin{table}[!t]
\renewcommand{\arraystretch}{1.15}
\caption{Device Coverage and Data Sufficiency in the Loaded Registry, 12 October 2026}
\label{tab:coverage}
\centering
\footnotesize
\setlength{\tabcolsep}{4pt}
\begin{tabular}{@{}l r@{}}
\toprule
Measure & Value \\
\midrule
\multicolumn{2}{@{}l}{\emph{Hourly summaries with Data Sufficiency = sufficient}} \\
Heart rate (485,951 hours) & 99.4\% \\
Respiratory rate (487,770 hours) & 99.1\% \\
Blood oxygen saturation (73,001 hours) & 89.5\% \\
\midrule
\multicolumn{2}{@{}l}{\emph{Valid readings per hourly summary, median (IQR)}} \\
Heart rate & 45 (36 to 52) \\
Respiratory rate & 53 (41 to 59) \\
Blood oxygen saturation & 51 (28 to 59) \\
\midrule
\multicolumn{2}{@{}l}{\emph{Hours with a reading outside the reference range}} \\
Heart rate, outside 60 to 100 & 58.5\% \\
Respiratory rate, outside 12 to 20 & 35.1\% \\
Blood oxygen saturation, below 95 & 95.2\% \\
Blood oxygen saturation, below 90 & 68.9\% \\
\midrule
\multicolumn{2}{@{}l}{\emph{Glucose monitoring (2,245 participants)}} \\
Sensor wear of 11 days & 85.3\% \\
Sensor wear of 10 days & 8.4\% \\
Sensor wear of 9 days or fewer & 5.8\% \\
Sensor wear of 12 days or more (14 or more: 2) & 0.5\% \\
Participants with $\geq$7 days & 98.3\% \\
Participants above 70\% hourly completeness & 98.9\% \\
\midrule
\multicolumn{2}{@{}l}{\emph{Wearable heart rate (1,999 participants)}} \\
Hours with a summary, median (IQR) & 251 (235 to 265) \\
Days with a summary, median (IQR) & 12 (12 to 13) \\
\midrule
\multicolumn{2}{@{}l}{\emph{Indoor environmental sensor (2,154 participants)}} \\
Duration, median (IQR) days & 10.5 (9.7 to 11.0) \\
Participants with $\geq$7 days & 91.6\% \\
\midrule
Participants with $\geq$7 days on all three streams & 1,709 of 2,280 (75.0\%) \\
\bottomrule
\end{tabular}
\vspace{1ex}
\begin{flushleft}
\scriptsize Sufficiency is the project definition of Section~\ref{sec:design}: three or more valid readings in the hour. Readings per hour are counted from the source files after the import's filters, over the same hours as the stored summaries.
Completeness is hours with a glucose summary divided by hours in the span
from first to last glucose day. Hours and days with a heart rate summary are counted per participant over the whole wear period, and days are UTC calendar days (Section~\ref{sec:granularity}). IQR is the interquartile range.
\end{flushleft}
\end{table}

\subsection{RQ3: Correctness}
\label{sec:rq3}

Verification compared every stored field of every event in the 20 non-core stages with the source, 64,850,380 values in all (Table~\ref{tab:verify}). Every value agreed within the tolerance of 0.01, no registry event lacked a source counterpart, and no source hour or record lacked an event. Of the agreements, 3,371,890 were pairs empty on both sides where a rule on the source requires it. Most were lists of flagged times for hours in which no reading was flagged. The rest were statistics that the source leaves undefined, such as the standard deviation of an hour with a single reading or the mean of a glucose hour holding only markers; the 40 OCTA acquisitions whose source has no superficial-plexus image; and 4,429 electrocardiogram header fields that the recording does not carry.

This result followed four repairs, each of a defect that only exhaustive, field-level verification revealed (\suppref{C}): a duplicated electrocardiogram event, 2,229 photographs without previews, 169,969 sleep events produced by segments that the export repeats verbatim or that end before they start, and 129,916 humidity hours whose list of flagged times had not been stored. None was a stored value that differed from its source. Each was a record missing, doubled or carried over from an artifact of the source, the kind of failure that the earlier sampled and count-only checks had not seen. A further 44,616 humidity lists of more than 20 flagged times had been stored in full by a load that predates the 20-entry cap; they agree with the source and were left as they are. The independent recomputation of glucose covered 46,421 hours, with every hour matched on both sides. All 464,210 values agreed, and the stored standard deviation differed from iglu's only by the divisor, $\sqrt{(n-1)/n}$.

Loading also surfaced properties of the source that any consumer of AI-READI should expect (Table~\ref{tab:dq}), classified by the harmonized data quality framework~\cite{kahn2016}. The catalog complements the release's own run of the OHDSI Data Quality Dashboard~\cite{blacketer2021} on its OMOP tables, which failed 49 of 1,241 checks. The phenomena that matter most for an ingestion layer are device placeholder values and non-finite numbers in the wearable exports, glucose markers that carry information and should not be discarded as noise, vendor-specific DICOM encodings, and sleep segments repeated verbatim in the export, which overstate summed sleep time by 43~percent for any consumer that does not remove them.

\begin{table}[!t]
\renewcommand{\arraystretch}{1.2}
\caption{Data Quality Phenomena of the Source, Classified by the Harmonized Framework~\cite{kahn2016}}
\label{tab:dq}
\centering
\scriptsize
\setlength{\tabcolsep}{3pt}
\begin{tabular}{@{}p{3.4cm} p{1.5cm} p{3.3cm}@{}}
\toprule
Phenomenon & Dimension & Handling \\
\midrule
Not-a-number values in the blood oxygen export (77 of 93 files sampled) & Completeness & Dropped before aggregation; finiteness guard on every value \\
Negative marker values in respiratory rate and stress (about half of readings); zero fill in heart rate & Plausibility & Dropped before aggregation \\
Glucose readings outside the sensor range exported as \code{High} and \code{Low} (13 of 60 files sampled) & Conformance (value type) & Counted as markers; excluded from numeric statistics, included in the time-in-range denominator \\
Nitrogen oxide coverage varies by participant; one has no usable value & Completeness & Hour omitted when the channel has no value; the stage is ten events short \\
\code{condition\_source\_value} truncated at 49 characters (3,388 of 12,375 rows) & Conformance (value) & Imported as is and documented \\
Directory tree nests each modality one level deeper than the manifest paths imply & Conformance (structure) & Path resolver accepts both layouts \\
No visit date in the registry master file & Completeness & Dates taken from the participant table; a record with no date is skipped, and no date is imputed \\
Six demographic attributes identify only 1,418 of 2,280 participants uniquely & Plausibility (uniqueness) & Join refused on ambiguous rows; new imports set the identifier directly \\
Time zone not stated in any export & Conformance (temporal) & Confirmed UTC by diurnal check; all timestamps UTC \\
Vendor-specific DICOM transfer syntax and YBR color space & Conformance (encoding) & Decoders declared; color space converted; 42,054 previews regenerated \\
JPEG~2000 pixel data wrapped in a JP2 file header, which DICOM does not permit (2,229 of 4,506 infrared images from one camera model) & Conformance (encoding) & Rejected by the first decoder, leaving no preview; found by verification, decoded with OpenJPEG, 2,229 previews regenerated \\
Sleep segments repeated verbatim in the wearable export (169,960 of 565,959, in 1,567 of 2,184 participants), overstating summed sleep time by 43\% & Plausibility (uniqueness) & Exact repeats dropped by the loader; repeats already loaded deleted \\
Sleep segments that end before they start (9) & Plausibility (temporal) & Dropped; events from an early load deleted \\
VOC index of 0 at sensor start-up, outside the documented 1 to 500 (0.009\% of readings) & Plausibility & Kept; the effect is negligible \\
Humidity documented as 0 to 1 but recorded in percent & Conformance (value) & Stored in percent, thresholds in percent \\
Superficial plexus en-face image reported as absent (40 of 24,560 angiography acquisitions) & Completeness & Loaded without a preview \\
Sex, race and ethnicity withheld from the public release; DICOM Patient Sex set to \code{M} in every file examined & Completeness; plausibility & Not loaded; corrected QT interval not classified \\
\bottomrule
\end{tabular}
\end{table}

\section{Discussion}
\label{sec:discussion}

\subsection{Principal findings}

\emph{RQ1.} An unmodified Tracker program can hold continuous glucose, wearable, environmental, electrocardiogram, diagnostic and imaging data for the same participants. By default, however, it provides no way to read a dense stream: the Capture application displays the events without summarizing them, so a usable view needs configured list columns, event reports or analytics.

\emph{RQ2.} The row-per-event model forces a choice of event unit, and at consumer-device density that unit cannot be the raw reading. The hourly summary makes the registry loadable, keeps the consensus glucose measures natural and aligns modalities on one clock. Its costs are the loss of the raw signal and, at this scale, dependence on analytics generation, which failed for the event table on this build.

\emph{RQ3.} The audited defects concerned the handling of failure; none was an error of arithmetic. Five controls specific to Tracker clients closed them. Exhaustive verification then found the remaining defects, which were defects of completeness that sampling and count-only checks had missed. Where a stage is small enough to check in full, it should be checked in full, and a check that shares the loader's code needs an independent counterpart.

\subsection{Relation to prior work}

As far as we can establish, this is the first report in the peer-reviewed literature of person-level physiological and environmental sensor streams held in DHIS2 Tracker. Device data have reached DHIS2 for cold-chain equipment, and images have been loaded into Tracker by script, both in community reports. The AI-READI consortium has mapped parts of the dataset into OMOP~\cite{ohdsiaireadi2024}, so the claim for AI-READI is only that this is its first loading into a health information system. At 8.09~million events the registry is large for research use of the platform and modest beside national production instances. Its reusable parts are implementation guidance: the data model, the loading lessons and the data quality catalog.

\subsection{Implications for diabetes care}

The registry places the consensus glucose measures~\cite{battelino2019, ada2026} in the same record as diagnoses, medications, laboratory results, indoor air, retinal images and electrocardiograms. Continuous glucose monitoring helps only if someone reviews the data~\cite{jancev2024}. In a DHIS2 deployment that reviewer can be a community health worker, whose interventions lower glycated hemoglobin in pooled trials~\cite{evans2025}, provided the data are presented in a form they can read. Indoor exposure aligned hourly with glucose makes within-person exposure-response analysis possible~\cite{yang2020}; retinal previews support screening in the same program~\cite{gulshan2016, wagner2020}; and the electrocardiogram stage could classify the corrected QT interval, a predictor of mortality in type~2 diabetes~\cite{cox2014}, once sex is recorded. The argument that a registry on the platform a ministry already runs is affordable~\cite{dhis2inaction} holds for the server tier. It does not hold for the devices, which are costly and little available in most low- and middle-income countries, and a separate research or program instance is a more realistic home for millions of sensor events than a production one. None of this has been shown to change a clinical outcome.

\subsection{From a loaded registry to a live platform}
\label{sec:live}

Hourly summaries are too coarse for decision support that forecasts glucose from five-minute readings~\cite{liu2023}, so a live deployment would be tiered: raw readings in a store built for them~\cite{ranjan2019}, summaries in Tracker, and alerts as events where health workers see them. This is the arrangement MIMIC uses for waveforms~\cite{johnson2016}. The tiers this work does not demonstrate are a patient-facing application, live device integration, interoperability through FHIR~\cite{abedian2025} or OMOP, validation rules and decision support. DHIS2 can now generate FHIR implementation guides from Tracker metadata~\cite{dhis2fhirig}, and standard diabetes and hypertension Tracker modules are being built for national programs~\cite{hispwdf2025}; aligning the core stages with them is the most direct route to adoption. The loading lessons are prerequisites for any continuous ingestion, because a streaming integration retries by nature.

\subsection{Limitations}

Several limitations concern the source data, the hourly model, the test environment and clinical readiness. Structural OCT and fluorescence lifetime imaging were not loaded, and the public release withholds sex, race and ethnicity. Within the hourly model, aggregation was checked independently only for glucose, activity proration smooths bursts within long segments, and storing timestamps in UTC displays UTC clock times in the Capture application. On the systems side, the server ran a pre-release build at default database settings, so its behaviors, including the analytics failure, should be confirmed on a released and tuned installation; per-modality load times were not recorded, loader throughput comes from a pilot, and the exported metadata package has been checked for completeness but not yet imported into a clean instance. Finally, the coverage figures describe research volunteers over about ten days, with glucose wear falling short of the 14 days the consensus recommends~\cite{battelino2019}; the reference-range flags fire in most hours and are not yet useful for review, and no clinician or health worker has used the registry.

\section{Conclusion}

An unmodified DHIS2 Tracker program held a multimodal type~2 diabetes registry of 2,280 participants and 8.09~million events loaded from AI-READI (RQ1). Its row-per-event model forced hourly summaries as the unit of sensor data. This made the registry loadable and aligned its modalities, but it leaves the raw signal to a separate tier and makes analytics generation the step that remains to be proven at scale (RQ2). Five lessons for Tracker clients made failed writes visible, and field-by-field verification against the source, with an independent recomputation of glucose, showed what the registry holds and found what earlier checks had missed (RQ3). Mobile capture, live device feeds, interoperability and decision support remain to be built before the registry could support care.


\section*{Declarations}

\emph{Ethics.} The study used only the de-identified public tier of the AI-READI dataset, under the data use agreement and license of the AI-READI Consortium~\cite{aireadilicense}; no further institutional review was sought. Every person with an account on the registry server holds AI-READI data access permission.

\emph{Funding.} This work received no specific funding and was carried out independently of the AI-READI Consortium.

\emph{Competing interests.} The authors declare no competing interests.

\emph{Data availability.} The AI-READI dataset is available from FAIRhub under its data use license~\cite{aireadi2025}; the license does not permit redistribution, so the loaded registry cannot be shared. Aggregate results reported here are in the code repository.

\emph{Code availability.} The pipeline, the exported program metadata and the verification scripts are available at \url{https://github.com/iupui-soic/dhis2-diabetes-registry}.

\emph{Supplementary material.} Appendices A to D (evaluation environment, interface behaviors and audit findings, verification details, and implementation notes) are provided as a separate document.

\section*{Acknowledgment}

The data used here are version~3.0.0 of the Flagship Dataset of Type~2 Diabetes from the AI-READI Project, which was produced by the AI-READI Consortium with support from the National Institutes of Health Bridge2AI Common Fund program under award OT2OD032644. The dataset was accessed under the AI-READI data use license, version~2.0, which limits use to research on type~2 diabetes and prohibits redistribution of the data. In keeping with that license, no participant-level values appear in this paper, and the accompanying code repository contains no data and no notebook output. The pipeline code, the metadata it creates and the verification scripts are available at
\url{https://github.com/iupui-soic/dhis2-diabetes-registry}. The authors thank the AI-READI Consortium for making the
dataset available, and the DHIS2 core team and community at the University of Oslo for the platform on which the registry is built.

\bibliographystyle{IEEEtran}
\bibliography{refs}

\newpage
\begin{IEEEbiographynophoto}{Saptarshi Purkayastha}
is an Associate Professor and Program Director of Health Informatics in the Department of Biomedical Engineering and Informatics at the Luddy School of Informatics, Computing, and Engineering, Indiana University Indianapolis, where he leads the Purkayastha Lab for Health Innovation. His research concerns electronic health record systems, including DHIS2 and OpenMRS, clinical decision support systems, and the integration of digital health data for population health.
\end{IEEEbiographynophoto}

\begin{IEEEbiographynophoto}{Aiswarya Nandan Perumbilly}
is a graduate student in the Master of Science in Health Informatics program at the Luddy School of Informatics, Computing, and Engineering, Indiana University Indianapolis, and a graduate research assistant in the Purkayastha Lab for Health Innovation. Her work concerns the configuration of DHIS2 Tracker programs and the loading and verification of multimodal health data.
\end{IEEEbiographynophoto}

\begin{IEEEbiographynophoto}{Zhenan Yin}
is a Ph.D. student in Biomedical and Health Informatics at the Luddy School of Informatics, Computing, and Engineering, Indiana University Indianapolis, and a graduate research assistant in the Purkayastha Lab for Health Innovation. He received the B.S. degree in electrical engineering from the University of California, San Diego, and worked for five years as a healthcare information technology project manager before beginning doctoral study. His research interests include multimodal learning, medical image analysis, clinical data mining and clinical artificial intelligence.
\end{IEEEbiographynophoto}

\begin{IEEEbiographynophoto}{Sherri L. Bucher}
is an Associate Professor and Chair of the Department of Community and Global Health at the Richard M. Fairbanks School of Public Health, Indiana University Indianapolis, with an adjunct appointment in Pediatrics at the Indiana University School of Medicine. She received the Ph.D. degree in child and developmental psychology from the University of Connecticut. Her research concerns maternal, newborn and child health, implementation science and digital health in resource-limited settings, including the mobile Helping Babies Survive platform built on DHIS2, and she received the 2024 Bantz-Petronio Translating Research into Practice Award.
\end{IEEEbiographynophoto}

\end{document}